\documentclass{article}

\usepackage{PRIMEarxiv}

\usepackage[utf8]{inputenc} 
\usepackage[T1]{fontenc}    
\usepackage{hyperref}       
\usepackage{url}            
\usepackage{booktabs}       
\usepackage{amsfonts}       
\usepackage{nicefrac}       
\usepackage{microtype}      
\usepackage{lipsum}
\usepackage{fancyhdr}       
\usepackage{graphicx}       
\usepackage{enumitem}
\usepackage{amsmath}
\usepackage{xcolor}
\usepackage{pifont}
\newcommand{\cmark}{\textcolor{green!70!black}{\ding{51}}}
\newcommand{\xmark}{\textcolor{red!70!black}{\ding{55}}}
\newcommand{\tmark}{\textcolor{orange!80!black}{$\triangle$}}
\graphicspath{{media/}}     

\title{SIGMAS: Second-Order Interaction-based Grouping for Overlapping Multi-Agent Swarms}

\author{
  Minah Lee \\
  The University of Texas at Dallas \\
  Richardson, TX, USA\\
  \texttt{minah.lee@utdallas.edu} \\
   \And
  Saibal Mukhopadhyay \\
  Georgia Institute of Technology \\
  Atlanta, GA, USA\\
  \texttt{saibal.mukhopadhyay@ece.gatech.edu} \\
}

\begin{document}
\maketitle

\begin{abstract}

Swarming systems, such as drone fleets and robotic teams, exhibit complex dynamics driven by both individual behaviors and emergent group-level interactions. Unlike traditional multi-agent domains such as pedestrian crowds or traffic systems, swarms typically consist of \textit{a few large} groups with \textit{inherent and persistent} memberships, making group identification essential for understanding fine-grained behavior. We introduce the novel task of \textbf{group prediction in overlapping multi-agent swarms}, where latent group structures must be inferred directly from agent trajectories without ground-truth supervision. To address this challenge, we propose \textbf{SIGMAS} (Second-order Interaction-based Grouping for Multi-Agent Swarms), a self-supervised framework that goes beyond direct pairwise interactions and model \textit{second-order interaction} across agents. By capturing how similarly agents interact with others, SIGMAS enables robust group inference and adaptively balances individual and collective dynamics through a learnable gating mechanism for joint reasoning. Experiments across diverse synthetic swarm scenarios demonstrate that SIGMAS accurately recovers latent group structures and remains robust under simultaneously overlapping swarm dynamics, establishing both a new benchmark task and a principled modeling framework for swarm understanding.

\end{abstract}

\section{Introduction}





\begin{figure*}[tb]
\centering
\includegraphics[width=0.85\linewidth]{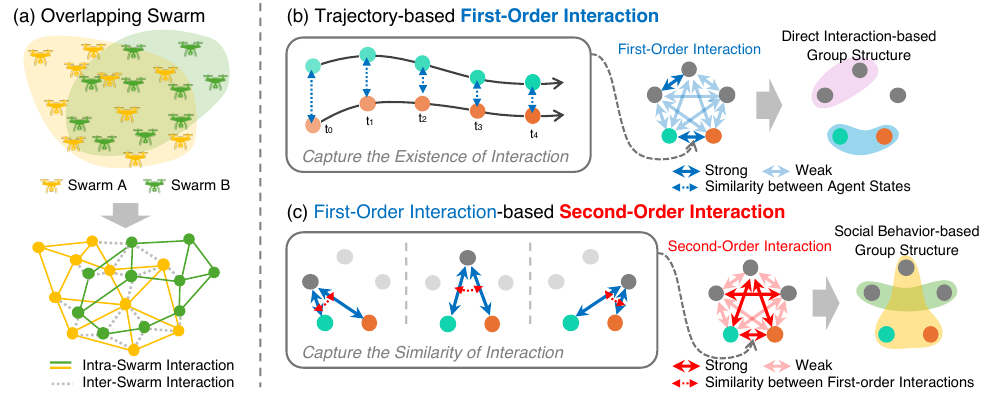}
\caption{(a) Overlapping swarms with distinct coordination patterns, making group identification difficult based on proximity alone. (b) \textit{First-order interactions} capture direct motion correlations between agents. (c) \textit{Second-order interactions} compare agents’ first-order attention patterns to measure how similarly they interact with others. This captures \textit{social behavior similarity} and enables robust group identification, even in the absence of direct interactions.} 
\label{fig:intro}
\end{figure*}

Swarming systems, including drone fleets, robotic teams, and autonomous vehicles, are key enablers of scalable and robust autonomous operations~\cite{yildirim2019decision, almalki2024synthesis,vlasceanu2019aerial, ribeiro2021multi, boubin2021programming}. Through simple local interactions, these systems exhibit emergent group-level behaviors that support coordination, collision avoidance, and efficient resource utilization in dynamic and uncertain environments~\cite{pretto2021building, mahmoudzadeh2022exploiting}. Compared to traditional multi-agent scenarios such as pedestrian motion, sports analytics, or traffic navigation~\cite{norambuena2023study, xu2023auxiliary, wong2024socialcircle}, swarming systems differ in two fundamental aspects. First, traditional settings typically involve many small groups with sparse interactions, whereas swarms typically consist of \textit{a few large groups} with dense intra-group coordination and complex inter-group dynamics, as illustrated in Figure~\ref{fig:intro}(a). Second, the notion of a “\textit{group}” also differs fundamentally. In traditional domains, groups are often defined by transient interactions arising from spatial proximity, while in swarming systems, group membership is an \textit{intrinsic and persistent} property of each agent. Agents from different swarms may even occupy the same spatial region at the same time, yet maintain distinct coordination patterns. 

A key challenge in multi-agent swarms is understanding the latent group structure that governs agent coordination. This is critical for robust behavior modeling and fine-grained interpretation of intra-swarm dynamics. To this end, we introduce the novel task of \textbf{group prediction in overlapping multi-agent swarms}, where the goal is to infer latent group memberships directly from agent trajectories without any ground-truth supervision. This self-supervised formulation is motivated by real-world swarm scenarios, where group affiliations are inherent but unobservable and may evolve over time. Since manually annotated group labels are rarely available in practice, we aim to learn group structure directly from motion dynamics without relying on explicit labels.

However, most existing trajectory prediction models are not well suited to this regime for two main reasons. First, many are designed for low-density scenes with sparse interactions, such as the ETH/UCY datasets~\cite{lerner2007crowds, pellegrini2009you}, which typically include only 1–6 agents within each observation and prediction window. Second, and more importantly, these models primarily rely on spatial proximity or pairwise motion similarity to model direct influence between agents, here referred to as \textit{first-order attention} mechanisms (Figure~\ref{fig:intro}(b)). While such mechanisms can be effective for short-term prediction, they are often insufficient to capture true group affiliations, especially in scenarios where multiple swarms overlap and occupy the same spatial region simultaneously, or where members of the same swarm are spatially dispersed. These limitations highlight the need for a specialized approach capable of understanding latent group structure in dense and dynamic swarm settings.

Therefore, we propose modeling social behavior through \textbf{second-order interactions}, where second-order interactions refer to the similarity between agents’ first-order attention patterns. In other words, instead of asking \textit{whether} two agents directly interact, we examine \textit{how similarly} they interact with others, as illustrated in Figure~\ref{fig:intro}(c). While first-order attention captures localized dependencies and can be effective for short-term trajectory forecasting, it often conflates transient spatial proximity with true group structure. In contrast, second-order interactions compare each agent’s full interaction profile, revealing shared behavioral patterns and uncovering latent affiliations. This approach is particularly useful in dense, overlapping, or partially converged swarm formations, where direct interactions are no longer a reliable signal of group membership.

To address these challenges, we introduce \textbf{SIGMAS (Second-order Interaction-based Grouping for Multi-Agent Swarms)}, a self-supervised framework for latent group inference in swarming systems. SIGMAS learns directly from agent trajectories without requiring manual annotations. SIGMAS features a novel swarm-level encoder that computes \textit{second-order interactions} by comparing agents’ first-order attention patterns. This enables SIGMAS to model interaction similarity and recover cohesive group structures even when agents are spatially entangled or dynamically dispersed. In addition, SIGMAS integrates a dynamic balancing mechanism to fuse individual and group-level dynamics, and applies spectral clustering for self-supervised group inference. These components jointly allow SIGMAS to capture both localized motion and global coordination patterns, enhancing group understanding in complex multi-agent environments.

This paper makes the following unique contributions:
\begin{itemize}
\item 
We formulate a novel task of \textbf{group prediction in overlapping multi-agent swarms}, where latent group memberships must be inferred from dense agent trajectories without ground-truth supervision. To the best of our knowledge, this is the first work to explicitly address this problem setting.

\item We present \textbf{SIGMAS}, a self-supervised framework that models group-level dynamics via second-order interaction modeling and adaptively fuses them with agent-level features through a learnable gating mechanism, enabling joint reasoning over local and global behaviors.

\item We empirically demonstrate that SIGMAS accurately recovers latent group structures and maintains robust performance across diverse and overlapping swarm scenarios.

\end{itemize}

\section{Background}

\subsection{Swarm Behavior and Interaction Modeling}

\begin{figure}[tb]
\centering
\includegraphics[width=0.5\linewidth]{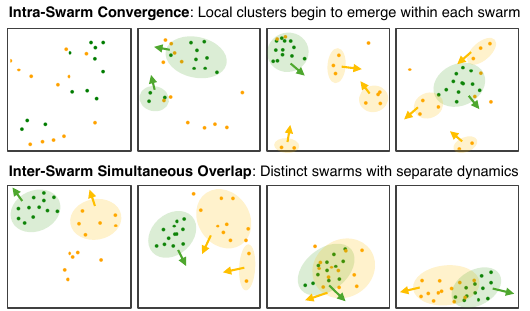}
\caption{Illustration of emergent swarm behaviors. \textbf{Top:} Intra-swarm convergence, where agents within the same swarm gradually align and form cohesive clusters. \textbf{Bottom:} Inter-swarm overlap, where distinct swarms simultaneously share spatial regions while maintaining separate coordination dynamics.} 
\label{fig:swarm_sample}
\end{figure}

Craig W. Reynolds~\cite{reynolds1987flocks} introduced three canonical rules for flocking behavior in agent-based systems: \textit{alignment}, \textit{cohesion}, and \textit{separation}. Each agent adjusts its velocity based on its local neighbors, leading to emergent collective behaviors such as coordinated motion and dynamic formation. We simulate these behaviors using the \texttt{agentpy} framework~\cite{foramitti2021agentpy}, which provides fine-grained control over agent-level parameters and swarm configurations. 
In our setup, agents within the same swarm share identical alignment, cohesion, and separation strengths, reflecting the assumption that \textit{group membership dictates underlying motion dynamics}. For agents in different swarms, only the separation rule applies to avoid collision. Figure~\ref{fig:swarm_sample} illustrates two key swarm phenomena that emerge from these local interaction rules:
\vspace{-0.1in}
\paragraph{\textbf{Intra-Swarm Convergence (Top Row).}} 
Agents within the same swarm gradually align their velocities, forming cohesive motion patterns over time. As interactions accumulate, local clusters emerge and evolve into tightly coordinated formations. The rate of convergence depends on intra-group interaction strength: higher alignment and cohesion lead to faster convergence and denser formations (green agents), while lower values yield slower and more diffuse structures (yellow agents).
\vspace{-0.1in}
\paragraph{\textbf{Inter-Swarm Overlap (Bottom Row).}} 
Distinct swarms may occupy overlapping spatial regions simultaneously while maintaining separate coordination dynamics. Depending on the relative strengths of intra- and inter-swarm interactions, agents may either preserve their original affiliations or partially mix across swarm boundaries.

These examples illustrate the challenge of inferring latent group structures in swarming systems. Spatial proximity alone is often misleading, as agents from different swarms may overlap while following distinct coordination patterns. This motivates the design of SIGMAS, which infers group membership by modeling \textit{second-order interaction}, offering a more robust foundation for group prediction in complex, overlapping swarm environments.

\subsection{Trajectory-based Interaction Modeling in Multi-Agent Systems}


\begin{table}[tb]
\caption{Comparison of key capabilities for group prediction in overlapping multi-agent swarms. 'Traj. Pred.' denotes trajectory prediction and 'Inter. Type' refers to the modeled interaction type. \tmark indicates that group modeling is used internally in the latent space to support trajectory prediction, but is not evaluated as a standalone group prediction objective.}
\centering
\begin{tabular}{lccccc}
\toprule
\textbf{Models} & \textbf{Method} & \textbf{Traj.} & \textbf{Inter.} & \textbf{Group}    & \textbf{Group}     \\
                &                      &\textbf{Pred.}  & \textbf{Type}   & \textbf{Modeling} & \textbf{Inference} \\
\midrule
SocialGAN~\cite{gupta2018social}               & LSTM             & \cmark & 1st & \xmark & \xmark \\
Trajectron++~\cite{salzmann2020trajectron++}   & LSTM             & \cmark & 1st & \xmark & \xmark \\
AgentFormer~\cite{yuan2021agentformer}         & Transformer      & \cmark & 1st & \xmark & \xmark \\
Leapfrog~\cite{mao2023leapfrog}                & Diffusion        & \cmark & 1st & \xmark & \xmark \\
GroupNet~\cite{xu2022groupnet}                 & GNN              & \cmark & 1st & \tmark & \xmark \\
SocialCircle~\cite{wong2024socialcircle}       & Transformer      & \cmark & 1st & \tmark & \xmark \\
\midrule
\textbf{SIGMAS (Ours)}                         & Transformer & \cmark & 1st \& 2nd & \cmark & \cmark \\
\bottomrule
\end{tabular}
\vspace{-0.15in}
\label{tab:capability_comparison}
\end{table}

Trajectory prediction has long been a core problem in multi-agent systems, with applications ranging from pedestrian modeling to human-robot interaction and autonomous navigation~\cite{gupta2018social, salzmann2020trajectron++, yuan2021agentformer}. A key challenge lies in capturing complex inter-agent dependencies, that influence motion, coordination, and intent understanding. Early approaches relied on handcrafted social rules~\cite{alahi2016social} or pooled neighborhood encodings~\cite{lee2017desire}, while recent works adopt graph-based and attention-based neural architectures for more expressive interaction modeling.

Most existing works model inter-agent interactions through \textit{first-order mechanisms}, where interactions are captured directly at the node or edge level. Node-centric approaches, such as SocialGAN~\cite{gupta2018social}, AgentFormer~\cite{yuan2021agentformer}, and Leapfrog~\cite{mao2023leapfrog}, focus on how individual agents respond to their neighbors based on proximity, attention, or goal-conditioned latent representations. Edge-centric models, such as Neural Relational Inference (NRI)~\cite{kipf2018neural}, infer latent pairwise relations between agents by predicting the presence and type of edges that best explain observed dynamics. These methods are effective in low-density environments with moderate agent counts and clearly separated groups, such as pedestrian crowds or physical particle simulations.

To address the limitations of purely pairwise modeling, several methods incorporate group-level cues as auxiliary signals to improve prediction accuracy. For example, GroupNet~\cite{xu2022groupnet} optimizes a social affinity matrix to identify latent groups, while SocialCircle~\cite{wong2024socialcircle} enforces temporal group consistency to enhance trajectory forecasting. Other methods leverage group priors or apply graph partitioning to better structure the interaction space~\cite{lee2024mart, bae2022learning, kim2024higher}. However, these methods do not explicitly model group emergence as a central objective.

In contrast, our work targets a fundamentally different regime: inferring latent group structure in dense, overlapping swarms. In such settings, group identity is not a transient outcome of spatial proximity but an intrinsic, persistent property of each agent, emerging from how similarly they interact with others. This requires new modeling methods that go beyond direct pairwise influence. To this end, we introduce \textit{second-order interactions}, which measure the similarity between agents’ first-order attention patterns. This mechanism enables robust group inference in densely interacting, dynamically evolving swarms where direct interactions may be ambiguous or unreliable. As summarized in Table~\ref{tab:capability_comparison}, SIGMAS is the only framework that simultaneously supports trajectory prediction, second-order interaction reasoning, and self-supervised group inference, all of which are essential for understanding complex group behavior in multi-agent swarms.

\section{Problem Statement}

Since group labels and the number of groups are unavailable during training, we formulate overlapping swarm identification as a \textbf{joint task of trajectory prediction and latent group inference} in a multi-agent swarm. Intuitively, given only the past $H$-step trajectories of agents (as shown in Figure~\ref{fig:intro}(a)), the model must jointly predict their future motion over $T$ time steps and identify the underlying group structure, i.e., $\mathcal{G} = \{ \{yellows\} , \{greens\} \}$. 

To formalize this task, let $N$ agents be observed up to time $t \leq 0$. This joint state is denoted by $\mathbf{X}^t = (x_1^t, x_2^t, \dots, x_N^t)$, where $x_n^t \in \mathbb{R}^d$ is the position of agent $n$ at time $t$. The observation history up to time $0$ is $\mathbf{X} = (\mathbf{X}^{-H}, \dots, \mathbf{X}^{0})$, with temporal horizon $H$. For future time steps $t > 0$, the corresponding joint state is $\mathbf{Y}^t = (y_1^t, y_2^t, \dots, y_N^t)$, where $y_n^t \in \mathbb{R}^d$ denotes the future position of agent $n$. The full future trajectories over $T$ time steps is $\mathbf{Y} = (\mathbf{Y}^1, \dots, \mathbf{Y}^T)$. 

In addition to predicting $\mathbf{Y}$, the goal is to infer latent group memberships for all agents.  
Let $\mathcal{G} = \{g_1, \dots, g_C\}$ denote the set of groups, where the number of groups $C$ is unknown during training, and each $g_c \subseteq \{a_1, \dots, a_N\}$ represents the subset of agents assigned to group $c$.  

The overall objective is to jointly estimate both the future trajectories $\mathbf{Y}$ and the latent group structure $\mathcal{G}$ from past observations $\mathbf{X}$ in a self-supervised manner: $p_\theta(\mathbf{Y}, \mathcal{G} \mid \mathbf{X})$, where $\theta$ denotes the model parameters. This formulation enables joint modeling of motion dynamics and latent group organization without requiring ground-truth group labels during training. 

\section{SIGMAS}

\begin{figure*}[tb]
	\centering
	\vspace{-0.1in}
    \includegraphics[width=0.8\linewidth]{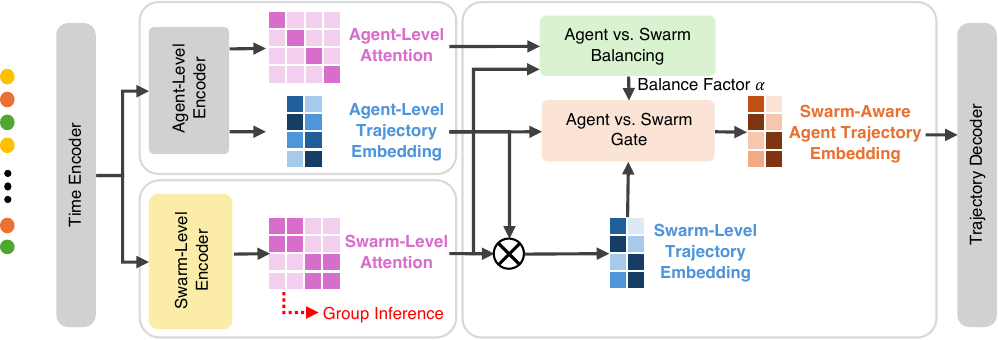}
    \vspace{-0.1in}
	\caption{SIGMAS architecture. Agent-level and swarm-level features are extracted via separate encoders, and fused using balancing and gating to form swarm-aware embeddings for trajectory prediction and group inference.} 
	\vspace{-0.1in}
	\label{fig:model}
\end{figure*}

We propose \textbf{SIGMAS (Second-order Interaction-based Grouping for Multi-Agent Swarms)}, a self-supervised framework that learns swarm-aware trajectory embeddings by jointly modeling individual motion and collective coordination. As illustrated in Figure~\ref{fig:model}, SIGMAS extracts agent-level and swarm-level trajectory embedding separately, and then fuses them to create a unified, swarm-aware representation for each agent. This fused embedding is used to predict future trajectories. In addition, the swarm-level attention matrix from the Swarm-Level Encoder is used to infer latent group structures during testing. The framework consists of three main components:  
(1) an \textbf{Agent-Level Encoder} that models first-order interactions and individual motion patterns,  
(2) a \textbf{Swarm-Level Encoder} that captures second-order interaction similarity across agents, and  
(3) a \textbf{Balancing \& Gating Module} that adaptively fuses individual and swarm-level representations.

\subsection{Agent-Level Encoder}
\label{sec:proposed_agent}

The Agent-Level Encoder captures localized dynamics and direct pairwise interactions among agents.  
Given temporally encoded trajectories, it produces two types of representations:  
\begin{itemize}[leftmargin=*]
    \item \textbf{Agent-Level Attention:} captures pairwise behavioral correlations (\textit{first-order interactions}) by computing attention scores based on trajectory similarity.
    \item \textbf{Agent-Level Trajectory Embedding:} represents each agent’s individual motion context.
\end{itemize}
We adopt the encoder architecture from AgentFormer~\cite{yuan2021agentformer}, which employs an agent-aware spatiotemporal attention mechanism. This allows each agent to attend differently to its neighbors depending on motion context, enabling flexible modeling of time-varying inter-agent dependencies. The resulting attention matrix reflects first-order interaction patterns, as illustrated in Figure~\ref{fig:intro}(b).

\subsection{Swarm-Level Encoder}
\label{sec:proposed_swarm}

\begin{figure*}[tb]
	\centering
	\vspace{-0.2in}
    \includegraphics[width=0.9\linewidth]{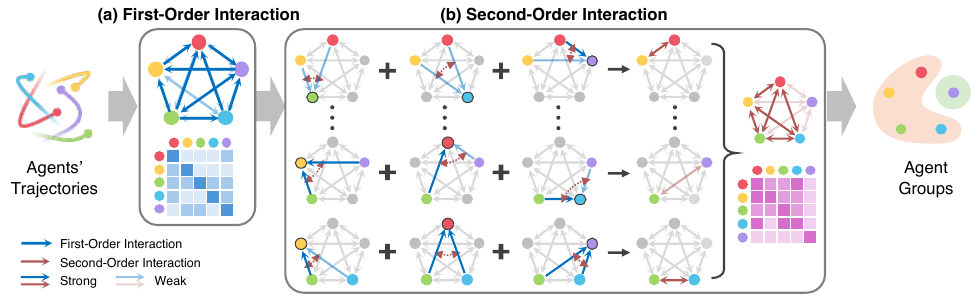}
    \vspace{-0.1in}
	\caption{Overview of second-order interaction modeling in the Swarm-Level Encoder. (a) First-order interactions encode directional influence between agents. (b) Second-order interactions compare first-order attention patterns across agents, revealing social behavior similarity. The resulting matrix highlights agents with similar interaction profiles, enabling group identification via clustering. In the attention matrix, rows correspond to attending agents and columns to agents being attended to. Darker cells indicate stronger outgoing attention. } 
	\vspace{-0.1in}
	\label{fig:swarm_encoder}
\end{figure*}

Swarm-Level Encoder captures \textit{social behavior similarity} among agents by modeling second-order interactions. It is motivated by the observation that agents within the same swarm tend to interact with others in similar ways, even without direct influence. The architecture is illustrated in Figure~\ref{fig:swarm_encoder}.

We first compute \textit{first-order interaction features} that capture direct correlations between agents based on their past trajectories. Given the trajectory embeddings $\mathbf{Q}, \mathbf{K} \in \mathbb{R}^{N \times d}$, we compute the first-order attention matrix as
\vspace{-0.1in}
\begin{align*}
\mathbf{A}^{(1)} = \operatorname{Softmax}\left( \frac{\mathbf{Q}\mathbf{K}^\top}{\sqrt{d}} \right)
\end{align*}
where $\mathbf{A}^{(1)}_{ij}$ represents how strongly agent $i$ attends to agent $j$. These attention scores reflect directional influence and localized behavioral dependencies among agents. 
For example, in Figure~\ref{fig:swarm_encoder}(a), the red agent attends strongly to the purple agent, while the green agent distributes its attention more broadly across the population.

To understand group-level structure, we compute \textit{second-order interactions} by comparing first-order attention patterns across agents:
\vspace{-0.1in}
\begin{align*}
\mathbf{A}^{(2)} = \mathbf{A}^{(1)} \mathbf{A}^{(1)\top}
\end{align*}
Each entry $\mathbf{A}^{(2)}_{ij}$ measures the similarity between agents $i$ and $j$ based on how similarly they attend to others, thus quantifying their \textit{social behavior similarity}. If two agents have similar attention distribution, they likely exhibit aligned social behaviors and therefore belong to the same swarm. In Figure~\ref{fig:swarm_encoder}(b), the green and blue agents both attend strongly to the red and purple agents, indicating similar social behavior and suggesting shared group affiliation.

The resulting second-order interaction matrix encodes pairwise similarity in social behavior. During training, we regularize this matrix to be \textit{symmetric} and \textit{positive semi-definite} (Section~\ref{sec:swarm_loss}), enforcing mutual consistency in social similarity. That is, if agent $i$ views agent $j$ as socially similar, the converse must also hold. During testing, $\mathbf{A}^{(2)}$ is passed to the clustering module (Section~\ref{sec:clustering}), which infers soft group assignments and, for instance, successfully groups the red, yellow, green, and blue agents into the same swarm.

\subsection{Agent vs. Swarm Balancing and Gating}
\label{sec:proposed_balancing}

As shown in Figure~\ref{fig:model}, the agent-level encoder produces two outputs: a first-order attention map $\mathbf{A}^{(1)} \in \mathbb{R}^{HN \times HN}$, which captures pairwise interactions across $H$ observed time steps, and an agent-level trajectory embedding $\mathbf{f}_n^{\text{agent}} \in \mathbb{R}^{HN \times d}$. The swarm-level encoder generates a second-order attention matrix $\mathbf{A}^{(2)} \in \mathbb{R}^{HN \times HN}$ that encodes second-order social similarity between agents. Using this matrix, we derive the swarm-level trajectory embedding via attention-weighted aggregation: $\mathbf{f}_n^{\text{group}} = \mathbf{A}^{(2)} \mathbf{f}_n^{\text{agent}} \in \mathbb{R}^{HN \times d}$.

The key challenge lies in how to combine $\mathbf{f}_n^{\text{agent}}$ and $\mathbf{f}_n^{\text{group}}$ into a unified representation that balances individual motion cues with group-level coordination. The optimal balance can vary depending on the swarm state, for example, the level of intra-group convergence, overlap across groups, or directional alignment, making static fusion suboptimal.

To address this, we introduce an adaptive gating mechanism that fuses the two embeddings in a context-aware manner. A soft gating coefficient $\alpha_n \in [0, 1]$ is computed for each agent using a multi-layer perceptron (MLP) over its agent-level embedding $\mathbf{f}_n^{\text{agent}}$, swarm-level embedding $\mathbf{f}_n^{\text{group}}$, and normalized group-assignment entropy $\mathcal{H}_n$:  
\[
\alpha_n = \text{MLP}\left([\mathbf{f}_n^{\text{agent}} \, \Vert \, \mathbf{f}_n^{\text{group}} \, \Vert \, \mathcal{H}_n]\right),
\]
where $\mathcal{H}_n = -\sum_{j=1}^{HN} A^{(2)}_{nj} \log A^{(2)}_{nj} / \log(HN)$ 
is the normalized entropy of the $n$-th row of $\mathbf{A}^{(2)}$. This entropy reflects the level of convergence within the swarm. High entropy indicates broad attention across peers (indicating a cohesive swarm), while low entropy suggests focused attention on a few agents (indicating pre-converged or fragmented structures).

The final trajectory embedding is computed as:
\[
\mathbf{f}_n^{\text{final}} = (1 - \alpha_n) \cdot \mathbf{f}_n^{\text{agent}} + \alpha_n \cdot \mathbf{f}_n^{\text{group}}.
\]
This dynamic fusion allows the model to adjust the contribution of individual versus group-level cues based on the agent’s social context.

\section{SIGMAS-based Prediction Framework}

\begin{figure}[tb]
	\centering
    \includegraphics[width=0.6\linewidth]{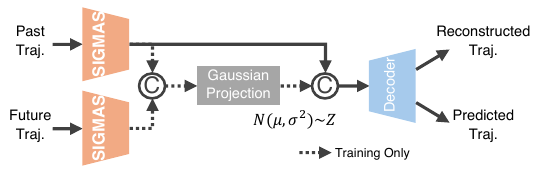}
	\caption{SIGMAS-based CVAE prediction framework. Dotted paths are used during training only.} 
	\label{fig:framework}
\end{figure}

We integrate SIGMAS into a conditional variational autoencoder (CVAE)-based multi-agent trajectory prediction framework following~\cite{xu2022groupnet} (Figure~\ref{fig:framework}). During the encoder stage, both past and future trajectories are used to estimate the posterior distribution parameters $(\mu_q, \sigma_q)$. SIGMAS is applied in this stage to produce swarm-aware trajectory embeddings that capture both individual and group-level behavior via first- and second-order interactions. 
In the decoder stage, a series of residual blocks take the past trajectory and a sampled latent embedding as input to generate both the predicted future trajectory and a reconstruction of the past. This residual design helps preserve input information while refining future predictions.

\subsection{Group Inference via Spectral Clustering.}
\label{sec:clustering}
The second-order attention matrix $\mathbf{A}^{(2)} \in \mathbb{R}^{N \times N}$ encodes meaningful group-relevant information by capturing similarities in social behavior. At early stages of swarm formation, when agents from the same group are still spatially dispersed, $\mathbf{A}^{(2)}$ may reflect multiple subgroup structures corresponding to partially converged clusters. In this case, a post-processing step during inference is necessary to merge these subgroups based on their dynamic similarity and given the number of groups $k$.

To infer group membership from $\mathbf{A}^{(2)}$, we treat it as a soft adjacency matrix and apply spectral clustering using the normalized graph Laplacian~\cite{ng2001spectral}. Specifically, we compute:
\[
\mathbf{L} = \mathbf{I} - \mathbf{D}^{-1/2} \mathbf{A}^{(2)} \mathbf{D}^{-1/2},
\]
where $\mathbf{D}$ is the degree matrix with diagonal entries $D_{ii} = \sum_j A^{(2)}_{ij}$.  
We then extract the $k$ smallest eigenvectors of $\mathbf{L}$ to obtain low-dimensional agent embeddings that capture the global interaction structure. $k$-means clustering is then applied to these embeddings to produce the predicted group assignments.

This procedure is fully self-supervised and relies on the geometry of the learned attention matrix to uncover cohesive swarm structures, even under spatial overlap or incomplete convergence, without requiring ground-truth labels.

\subsection{Training Objectives}
The training objective of SIGMAS consists of three components: a trajectory prediction loss, a swarm-level attention regularization loss, and a balancing factor regularization loss. These are jointly optimized to ensure accurate forecasting while promoting structured group-aware representations.

\paragraph{\underline{\textbf{Trajectory Prediction Loss}}}
We adopt a composite loss~\cite{xu2022groupnet} that includes: (1) an evidence lower bound (ELBO) loss from the CVAE formulation, (2) a reconstruction loss on the past trajectory, and (3) a diversity-promoting variety loss following Social-GAN:
\[
\mathcal{L}_{\text{elbo}} = 
\big\| \widehat{\mathbf{X}}^{+} - \mathbf{X}^{+} \big\|_2^2 
+ \, \mathrm{KL}\!\left( \mathcal{N}(\mu_q, \mathrm{Diag}(\sigma_q^2)) 
\,\|\, \mathcal{N}(0, \lambda I) \right), 
\]
\[
 \mathcal{L}_{\text{rec}} = 
\big\| \widehat{\mathbf{X}} - \mathbf{X} \big\|_2^2, 
\]
 \[
 \mathcal{L}_{\text{variety}} = 
\min_k \big\| \widehat{\mathbf{X}}^{+(k)} - \mathbf{X}^{+} \big\|_2^2, 
\]
 \[
 \mathcal{L}_{\text{traj}} = \mathcal{L}_{\text{elbo}} + \mathcal{L}_{\text{rec}} + \mathcal{L}_{\text{variety}},
\]
where $\widehat{\mathbf{X}}$ is the reconstructed past trajectory, $\widehat{\mathbf{X}}^{+}$ is the predicted future trajectory, $\mathbf{X}^{+}$ is the ground-truth future trajectory, and $\widehat{\mathbf{X}}^{+(k)}$ represents the $k$-th stochastic sample. $\|\cdot\|_2$ denotes the $\ell_2$ norm, and $\mathrm{KL}(\cdot\|\cdot)$ denotes the Kullback–Leibler divergence. This loss encourages the model to generate accurate, diverse predictions while preserving past dynamics and maintaining a regularized latent representation.

\paragraph{\underline{\textbf{Swarm-Level Attention Regularization}}}
\label{sec:swarm_loss}

To ensure that the second-order interaction matrix $\mathbf{A}^{(2)}$ captures meaningful group structure, we introduce a regularization loss on the swarm-level similarity matrix $\mathbf{S} \in \mathbb{R}^{N \times N}$, obtained by averaging $\mathbf{A}^{(2)}$ across the observation horizon. We enforce two key properties to make $\mathbf{S}$ a valid affinity kernel:

\begin{itemize}[leftmargin=*]
    \item \textbf{Symmetry Loss.} 
    Social similarity should be mutual: if agent $i$ is behaviorally similar to agent $j$, then the reverse should also hold. We eencourage symmetry by minimizing:
    \[
    \mathcal{L}_{\text{sym}} = \| \mathbf{S} - \mathbf{S}^\top \|_F^2,
    \]
    which penalizes asymmetric affinity scores and promotes bidirectional consistency across agent pairs.

    \item \textbf{Positive Semi-Definite (PSD) Loss.} 
    To ensure $\mathbf{S}$ is a valid inner-product kernel for spectral clustering, we encourage it to be positive semi-definite. Specifically, we penalize negative eigenvalues:
    \[
    \mathcal{L}_{\text{psd}} = \sum_{i=1}^{N} \left[ \max(0, -\lambda_i) \right]^2,
    \]
    where $\lambda_i$ is the $i$-th eigenvalue of $\mathbf{S}$. This regularization ensures geometric coherence in the learned affinity space.

\end{itemize}

\noindent The total swarm-level regularization is defined as:
\[
\mathcal{L}_{\text{swarm}} = \mathcal{L}_{\text{sym}} + \mathcal{L}_{\text{psd}}.
\]
These constraints guide $\mathbf{A}^{(2)}$ to form a symmetric, geometrically valid similarity matrix that supports interpretable and robust group identification during inference.

\paragraph{\underline{\textbf{Balancing Factor Regularization}}}

To enable adaptive fusion between agent-level and swarm-level dynamics, the model learns a gating factor $\alpha_n \in [0,1]$ for each agent $n$, as introduced in Section~\ref{sec:proposed_balancing}. Ideally, $\alpha_n$ should consider both the degree of swarm convergence as well as agent’s local motion.

To guide this behavior, we introduce a regularization loss that aligns the learned $\alpha_n$ with a target value based on the entropy of the swarm-level attention. Let $\mathcal{H}_n$ denote the normalized entropy of the $n$-th row of the second-order attention matrix $\mathbf{A}^{(2)}$. We define the target balancing factor as:
\[
\tilde{\alpha}_n = 1 - \mathcal{H}_n.
\]
Low entropy indicates that the agent strongly attends to a small subset of peers, suggesting sub-convergence and a stronger need to rely on swarm-level information. We encourage alignment via the loss:
\[
\mathcal{L}_{\text{balance}} = \frac{1}{N} \sum_{n=1}^N \left( \alpha_n - \tilde{\alpha}_n \right)^2.
\]
This term promotes dynamic adjustment of the agent’s attention balance based on the current level of social organization, improving adaptability across diverse swarm formations.

\paragraph{\underline{\textbf{Total Loss}}}
The total objective is a weighted sum of the three terms:\vspace{-0.1in}
\[
\mathcal{L}_{\text{total}} = \mathcal{L}_{\text{traj}} + \mathcal{L}_{\text{swarm}} + \mathcal{L}_{\text{balance}},
\]

\section{Simulation Results}

\subsection{Experimental Setup}

\paragraph{\textbf{Datasets}}
Due to the lack of public datasets for large, overlapping multi-agent swarms, we simulate swarm dynamics using a flocking-based agent simulator (AgentPy~\cite{foramitti2021agentpy}) based on Reynolds’ rules~\cite{reynolds1987flocks}. To constrain agent motion, we apply a reflective boundary condition and keep agents bouncing back at the simulation edge. The dataset contains 120 sequences in total: 100 for training, 10 for validation, and 10 for testing. Each sequence consists of 24 agents divided into two groups of 12 agents each, moving over 200 time steps. Agents are initialized with random positions and velocities and gradually converge into coherent swarms through local interactions.

\paragraph{\textbf{Evaluation Metrics}}



Let $\mathbf{z} = (z_1, \dots, z_N)$ denote the ground-truth group labels and $\hat{\mathbf{z}} = (\hat{z}_1, \dots, \hat{z}_N)$ the predicted assignments. We evaluate clustering performance using three standard metrics:

\noindent(1) \textit{Adjusted Rand Index (ARI)~\cite{hubert1985comparing} }: 
ARI measures pairwise agreement between predicted and true groupings, correcting for chance. Let $n_{ij}$ be the number of agents in true cluster $i$ assigned to predicted cluster $j$, $a_i = \sum_j n_{ij}$, $b_j = \sum_i n_{ij}$, and $N = \sum_{i,j} n_{ij}$. Then:
\[
\text{ARI} = 
\frac{\sum_{i,j} \binom{n_{ij}}{2} - 
\frac{\sum_i \binom{a_i}{2} \sum_j \binom{b_j}{2}}{\binom{N}{2}}}
{\frac{1}{2} \left[\sum_i \binom{a_i}{2} + \sum_j \binom{b_j}{2} \right] - 
\frac{\sum_i \binom{a_i}{2} \sum_j \binom{b_j}{2}}{\binom{N}{2}}}.
\]
ARI ranges from $-1$ (anti-correlation) to $1$ (perfect match), with $0$ indicating random labeling.

\noindent(2) \textit{Normalized Mutual Information (NMI)~\cite{strehl2002cluster} }:
NMI measures the mutual information between ground-truth and predicted clusters, normalized by their entropies. Let $p_{ij}$, $p_i$, and $p_j$ denote the joint and marginal probabilities from the contingency table:
\[I(\mathcal{C}; \hat{\mathcal{C}}) = \sum_{i,j} p_{ij} \log \left( \frac{p_{ij}}{p_i p_j} \right)\]
\[H(\mathcal{C}) = -\sum_i p_i \log p_i, \quad H(\hat{\mathcal{C}}) = -\sum_j p_j \log p_j\].
The normalized score is:\vspace{-0.1in}
\[
\text{NMI}(\mathcal{C}, \hat{\mathcal{C}}) = 
\frac{2 I(\mathcal{C}; \hat{\mathcal{C}})}{H(\mathcal{C}) + H(\hat{\mathcal{C}})}.
\]
which lies in $[0,1]$, with higher values indicating stronger agreement.

\noindent(3) \textit{F-score~\cite{powers2020evaluation} }:
The F-score combines pairwise precision and recall. For all agent pairs, define $TP$ (true positives), $FP$ (false positives), and $FN$ (false negatives). Then:
\[
\text{Precision} = \frac{TP}{TP + FP}, \quad 
\text{Recall} = \frac{TP}{TP + FN}, \quad
F = \frac{2 \cdot \text{Precision} \cdot \text{Recall}}{\text{Precision} + \text{Recall}}.
\]
F-score balances both false groupings and missed groupings, with $F \in [0,1]$.

\subsection{Validation on Group Inference}

We validate the group inference capability of SIGMAS in a simulated environment with two swarming groups exhibiting overlapping trajectories and distinct internal dynamics. While both groups follow Reynolds' flocking rules, they differ in cohesion, alignment, and movement speed, resulting in heterogeneous intra-swarm behavior. No explicit inter-group interaction is applied, creating scenarios where agents from different swarms spatially interleave. \texttt{Swarm A} in Table~\ref{tab:swarm_scenarios} is used.

\begin{figure}[tb]
	\centering
    \includegraphics[width=0.6\linewidth]{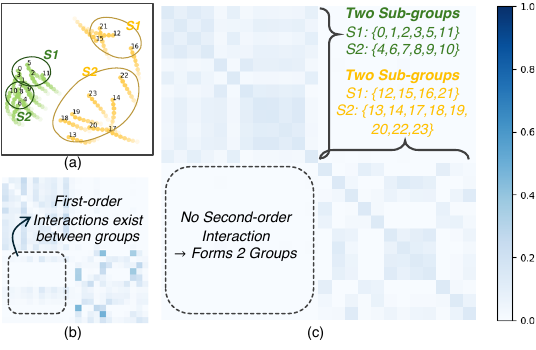}
	\caption{Comparison of first-order and second-order attention for group inference. (a) Agent trajectories show two overlapping swarms. (b) Agent-level attention $\mathbf{A}^{(1)}$ reflects direct interaction relevance for trajectory prediction. (c) Swarm-level attention $\mathbf{A}^{(2)}$ captures second-order interaction similarity across agents, enabling accurate latent group discovery via spectral clustering.} 
	\label{fig:clustering}
\end{figure}

\begin{table}[tb]
\caption{Swarm configurations for different scenarios.}
\vspace{-0.1in}
\centering
\begin{tabular}{lcccc}
\toprule
Swarm Config  & Cohesion & Separation & Alignment & Speed\\
\midrule
\texttt{Swarm A}   &  [0.005, 0.007] & [0.1, 0.2] & [0.3, 0.1] & [1.0, 1.2]\\
\texttt{Swarm B}   &  [0.005, 0.010] & [0.1, 0.3] & [0.1, 0.3] & [1.0, 1.3]\\
\texttt{Swarm C}   &  [0.005, 0.010] & [0.1, 0.4] & [0.1, 0.3] & [1.0, 1.5]\\
\bottomrule
\end{tabular}
\label{tab:swarm_scenarios}
\vspace{-0.2in}
\end{table}

\paragraph{\textbf{Latent Clustering}}

This experiment demonstrates how the proposed swarm-level encoder, based on second-order interactions, enables accurate group inference in partially converged, overlapping swarm scenarios. Figure~\ref{fig:clustering}(a) visualizes trajectories from the two swarms, where each swarm forms local sub-groups that have not yet fully converged. Effective group prediction in such scenarios requires identifying fine-grained sub-structures while preserving global swarm affiliation.

Figure~\ref{fig:clustering}(b) presents the first-order attention matrix $\mathbf{A}^{(1)}$ from the agent-level encoder (Section~\ref{sec:proposed_agent}), which captures direct behavioral correlations. While this matrix is useful for trajectory prediction, it primarily reflect transient spatial dependencies. For example, agents \{13, 18, 19\} from the yellow swarm attend to nearby green agents to avoid collisions. Such interactions are helpful for motion forecasting but do not reliably indicate shared group membership, making $\mathbf{A}^{(1)}$ less effective for clustering.

In contrast, Figure~\ref{fig:clustering}(c) presents the second-order attention matrix $\mathbf{A}^{(2)}$ (Section~\ref{sec:proposed_swarm}), which compares first-order attention patterns across agents to capture social behavior similarity. This matrix exhibits a clear block-diagonal structure with low inter-swarm similarity, effectively separating the green and yellow swarms. It also uncovers four latent sub-groups (S1--S4). Spectral clustering analyzes the eigenspace of the graph Laplacian and merges S1 and S2 into one swarm, and S3 and S4 into another, based on the known number of groups.

This example demonstrates that second-order interaction modeling captures both intra-swarm coordination and inter-swarm separation, enabling robust group inference even under partial convergence.

\paragraph{\textbf{Individual vs Swarm Balancing}}
Since SIGMAS is trained with trajectory supervision, the utility of latent group information depends on the swarm’s configuration. It is therefore crucial to adaptively balance agent-level and swarm-level embeddings to capture both individual motion and collective behavior. To achieve this, SIGMAS learns a dynamic gating factor $\alpha \in [0,1]$ that adjusts the contribution of swarm-level information relative to individual dynamics.

\begin{figure}[tb]
	\centering
    \includegraphics[width=0.6\linewidth]{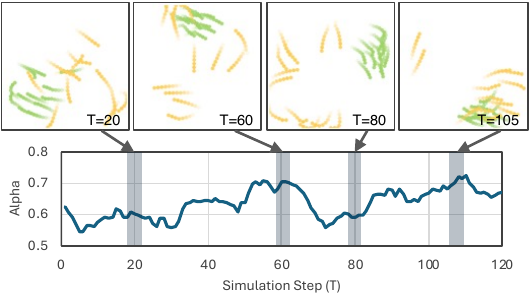}
	\caption{Adaptive balancing factor $\alpha$ over time in response to swarm dynamics. The model increases $\alpha$ during inter-swarm overlap and decreases it when individual motion cues are more predictive. $\alpha = 0$ corresponds to relying solely on agent-level trajectory embeddings, while $\alpha = 1$ reflects full reliance on swarm-level embeddings.} 
	\label{fig:alpha}
\end{figure}

Figure~\ref{fig:alpha} shows how $\alpha$ evolves over time in a multi-swarm simulation, along with corresponding snapshots at selected timesteps. Early in the sequence (e.g., $T=20$), agents are loosely coordinated, so trajectory prediction relies more on individual behavior, resulting in low $\alpha$. As intra-swarm cohesion increases, agents begin to exhibit more structured group motion. However, when swarms remain well-separated (e.g., $T=80$), group identity becomes trivial, and individual cues again dominate, lowering $\alpha$.

In contrast, during inter-swarm overlap (e.g., $T=60$, $T=105$), spatial proximity between agents from different swarms introduces ambiguity. In these cases, swarm-level context becomes essential for resolving latent affiliations, leading to higher $\alpha$ values. This adaptive gating mechanism enables SIGMAS to dynamically shift its reliance based on the swarm interactions.

\paragraph{\textbf{Comparison with Prior Work}}


\begin{table}[tb]
\caption{Comparison of group prediction performance across baselines.}
\label{tab:model_comparison}
\centering
\begin{tabular}{lccc}
\toprule
& AgentFormer~\cite{yuan2021agentformer} & SIGMAS\_IG & \textbf{SIGMAS} \\
\midrule
ARI$\uparrow$     & 0.3155  & 0.3306   &  \textbf{0.4045}\\
NMI$\uparrow$     & 0.3739  & 0.3860   &  \textbf{0.4086}\\
F-score$\uparrow$ & 0.6863  & 0.6887   &  \textbf{0.6968}\\
\bottomrule
\end{tabular}
\end{table}

\begin{figure}[tb]
	\centering
    \includegraphics[width=0.5\linewidth]{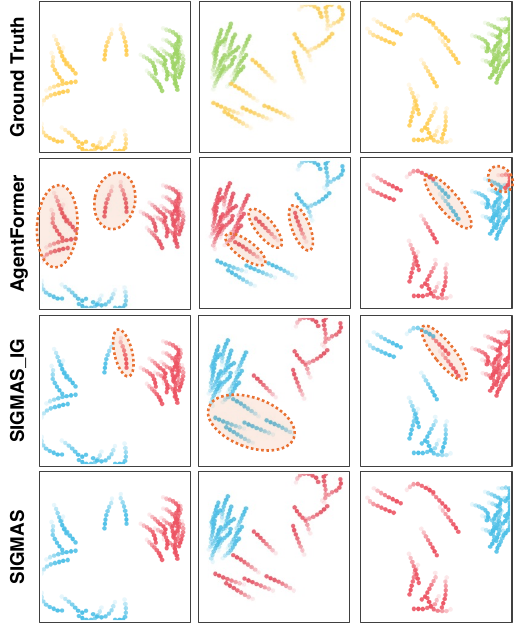}
	\caption{Qualitative results of group prediction. Red dotted ellipses indicate incorrect group assignments. Group labels are permutation-invariant, so different colors indicate distinct predicted clusters. Results videos are in Supplementary.} 
	\label{fig:pred_sample}
\end{figure}

We evaluate group prediction accuracy by comparing SIGMAS against relevant baselines. As group prediction in swarming systems is a novel task, few existing models are directly applicable. Furthermore, many multi-agent trajectory models are not designed to handle large-scale swarms. We consider the following baselines:
\begin{itemize}[leftmargin=*]
\item \textbf{AgentFormer~\cite{yuan2021agentformer}}: A transformer-based trajectory model that uses self-attention to capture pairwise interactions. For group prediction, we apply spectral clustering to its final attention matrix.
\item \textbf{SIGMAS\_IG}: A variant of our model that uses both encoders during training but applies spectral clustering to the agent-level attention matrix $\mathbf{A}^{(1)}$ instead of the second-order matrix.
\end{itemize}
Our full model, SIGMAS, performs group inference using $\mathbf{A}^{(2)}$ from the swarm-level encoder, which captures second-order social behavior similarity.

Figure~\ref{fig:pred_sample} presents qualitative results. AgentFormer and SIGMAS\_IG frequently produce incorrect groupings in cases with spatial overlap or partial convergence, often grouping agents with similar motion direction rather than true swarm affiliation (see red ellipses). In contrast, SIGMAS successfully recovers the underlying group structure by leveraging second-order interactions, enabling robust inference even under ambiguous conditions. These observations are reflected in Table~\ref{tab:model_comparison}, where SIGMAS achieves better performance across all clustering metrics.

\subsection{Case study: Intra-Swarm Heterogeneity}

\begin{figure}[tb]
	\centering
    \includegraphics[width=0.5\linewidth]{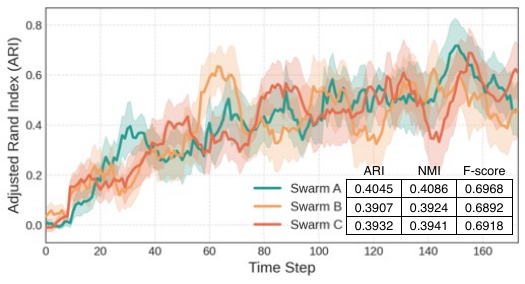}
	\caption{Group prediction performance across three heterogeneous swarm configurations (\texttt{Swarm A, B, C}) with increasing intra-swarm disparity. Shaded regions indicate standard deviation across multiple runs. Results videos are in Supplementary.} 
	\label{fig:intra_hetero}
\end{figure}

To evaluate the robustness of SIGMAS under varying swarm dynamics, we conduct experiments across three heterogeneous swarm settings (\texttt{Swarm A, B, C} in Table~\ref{tab:swarm_scenarios}). In each case, the two swarms differ in cohesion, separation, alignment, and speed, simulating different levels of swarm heterogeneity. A single SIGMAS model trained on \texttt{Swarm A} is evaluated on all three configurations to assess its generalizability.

Figure~\ref{fig:intra_hetero} shows the Adjusted Rand Index (ARI) over time for all three scenarios. As agents gradually converge into coherent swarms, ARI improves steadily. Importantly, SIGMAS achieves consistent performance across all settings despite the growing heterogeneity in swarm dynamics. This indicates that SIGMAS reliably infers group structure even under diverse intra-swarm dynamics. Similar trends are observed for NMI and F-score, confirming the model’s generalizability to a wide range of heterogeneous swarm behaviors.

\section{Conclusions}

Swarming systems exhibit complex, emergent behaviors driven by dense local interactions. Unlike traditional multi-agent settings, multiple swarms often operate in overlapping regions while maintaining persistent group identities, making latent group structure inference both critical and underexplored. In this paper, we introduce the novel task of \textbf{group prediction in overlapping multi-agent swarms}, where the objective is to recover group memberships directly from trajectory data without ground-truth supervision. This problem setting requires modeling beyond local motion interactions to capture global coordination behavior inherent in swarm dynamics

To address this challenge, we propose \textbf{SIGMAS}, a self-supervised framework that jointly captures individual and collective behavior through a novel \textit{second-order interaction} mechanism. By measuring how similarly agents attend to others, SIGMAS quantifies social behavior similarity and supports robust group inference even in spatially entangled or partially converged swarm formations. An adaptive gating module fuses agent-level and swarm-level embeddings based on social context. Experiments demonstrate that SIGMAS outperforms prior baselines and generalizes well across diverse heterogeneous swarm scenarios.

\section*{Acknowledgments}
This work is supported by the Defense Advanced Research Projects Agency (DARPA) under Grant Numbers GR00019153. The views and conclusions contained in this document are those of the authors and should not be interpreted as representing the official policies, either expressed or implied, of the Department of Defence, DARPA, or the U.S. Government.

\bibliographystyle{unsrt}  
\bibliography{ref}

\end{document}